\documentclass[12pt]{iopart}
\usepackage{iopams}
\usepackage{graphicx}  
\begin{document}

\title[Two-parameter generalization of the logarithm and entropy]{Two-parameter generalization of the logarithm and exponential functions and Boltzmann-Gibbs-Shannon entropy}

\author{Veit Schw\"ammle and Constantino Tsallis}

\address{Centro Brasileiro de Pesquisas Fisicas \\ 
Rua Xavier Sigaud 150, 22290-180 Rio de Janeiro-RJ, Brazil}
\ead{veit@cbpf.br, tsallis@cbpf.br}
\begin{abstract}
The $q$-sum $x \oplus_q y \equiv x+y+(1-q)\,xy$ ($x \oplus_1 y=x+y$) and the $q$-product
$x\otimes_q y \equiv [x^{1-q} +y^{1-q}-1]^{\frac{1}{1-q}}$ ($x\otimes_1 y=x\,y$) emerge
naturally within nonextensive statistical mechanics. We show here how they lead to
two-parameter (namely, $q$ and $q^\prime$) generalizations of the logarithmic and exponential functions (noted respectively
$\ln_{q,q^\prime}x$ and $e_{q,q^\prime}^{\,x}$), as well as of the Boltzmann-Gibbs-Shannon
entropy $S_{BGS}\equiv -k \sum_{i=1}^Wp_i \ln p_i$ (noted $S_{q,q^\prime}$). The remarkable
properties of the $(q,q^\prime)$-generalized logarithmic function make the entropic form
$S_{q,q^\prime} \equiv k \sum_{i=1}^W p_i \ln_{q,q^\prime}(1/p_i)$ to satisfy, for large
regions of $(q,q^\prime)$, important properties such as {\it expansibility},
{\it concavity} and {\it Lesche-stability}, but not necessarily {\it
  composability}.   
\end{abstract}



\section{INTRODUCTION}

Boltzmann--Gibbs ($BG$) statistical mechanics provides a powerful tool for
understanding how fast microscopic physics with short--range interactions
affects the thermodynamics of a system on much larger space--time scales. It
is based on the ({\it Boltzmann-Gibbs} or {\it Boltzmann-Gibbs-Shannon}) entropy 
\begin{equation}
S_{BGS}\equiv -k\sum_{i=1}^W p_i \ln p_i=k\sum_{i=1}^W p_i \ln \frac{1}{p_i}  \,\,\,\,(\sum_{i=1}^Wp_i=1)   \,,
\end{equation}
hence, if $p_i=1/W \,(\forall i)$, $S_{BGS}=k \ln W$. 

However, the $BG$ theory turns out to be inadequate for various complex
natural, artificial and social systems. Such is, for instance, the case where
a zero maximal Lyapunov exponent is present (e.g., long--range--interaction
many--body Hamiltonian systems, nonlinear dynamical systems at the edge of chaos, and
others). Typically, such situations are dominated  by power-laws instead of
exponential distributions \cite{next:04,next:05}. 
  In order to deal with at least part of such systems, it was proposed in 1988
\cite{Tsallis1988}, a generalization of the $BG$ theory, referred to as {\it nonextensive statistical mechanics} \cite{CuradoTsallis:91,TsallisMendes:98}. It is based on the entropy
\begin{equation}
S_{q}\equiv k\frac{1- \sum_{i=1}^W p_i^q}{q-1}=k\sum_{i=1}^W p_i \ln_q (\frac{1}{p_i}) \,\,\,\,(q \in {\cal R}, \sum_{i=1}^Wp_i=1)       \,,
\end{equation} 
(hence, if $p_i=1/W \,(\forall i)$, $S_q=k \ln_q W$) 
where
\begin{equation}
\ln_q x \equiv  \frac{x^{1-q}-1}{1-q} \,\,\,\,(\ln_1 x=\ln x) \,.
\end{equation} 
This entropy succeeds to widen the range of applicability of statistical mechanical concepts to many systems.
It fulfills thermodynamically relevant properties such as expansibility,
composability, Lesche--stability and
concavity (for $q>0$). The roles played in the traditional theory by the logarithmic and exponential functions are played, in the generalized theory, by
the so--called $q$-logarithm, as given by Eq. (3), and its inverse, the $q$-exponential \cite{Tsallis:94b}:
\begin{equation}
e_q^x \equiv [1+(1-q)x]_+^{\frac{1}{1-q}}  \;\;\;\;(e_1^x=e^x),
\end{equation} 
where $[...]_+$ vanishes if its argument is nonpositive.
This naturally led to the construction of a 
family of functions based on the $q$--generalized logarithm and its inverse
function~\cite{Borges:98b}.

Consistently, nonlinear generalized algebraic forms emerge (see also~\cite{Kaniadakis:02}), namely the {\it
  $q$--sum}~\cite{Borges:2004} and
the {\it $q$--product}~\cite{Borges:2004,Nivanen:2003} respectively defined as follows:
\begin{equation}
x\oplus_q y \equiv x+y + (1-q) xy \;\;\;\;(x\oplus_1 y=x+y) \,,
\end{equation}
and 
\begin{equation}
x\otimes_qy \equiv \left(x^{1-q}+y^{1-q} -1 \right)^{\frac{1}{1-q}}   \;\;\;\;(x\otimes_1y=xy)\,.
\end{equation}
Additionally, with these mathematical generalizations of the sum and the product, 
a new algebra could be constructed which appears to have some peculiar 
properties not found for other algebras~\cite{Borges:2004,Nivanen:2003}. 
The $q$-product and the
$q$-sum maintain the structure given by Eq.~(\ref{eq:q_rel1}) by
converting the sum and the product into these non--linear operators.

We can straightforwardly verify the following properties:
\begin{eqnarray}
  \label{eq:q_rel1}
\ln(x \cdot y) = \ln x + \ln y \\
  \label{eq:q_rel2}
\ln_q(x \cdot y) = \ln_q x \oplus_q \ln_q y\\
  \label{eq:q_rel3}
\ln_q(x \otimes_q y) = \ln_q x + \ln_q y
\end{eqnarray}
In the following section (Sec. 2) we show that an uniquely further generalized logarithmic function, denoted $\ln_{q,q^{\prime}} x$, exists which enables the unification of 
Eqs.~(\ref{eq:q_rel1}-\ref{eq:q_rel3}) as follows: 
\begin{equation}
\ln_{q,q^{\prime}}(x \otimes_q y) = \ln_{q,q^{\prime}} x \oplus_{q^{\prime}} \ln_{q,q^{\prime}} y \,.
  \label{eq:fqq1}
\end{equation}
In Sec. 3 we present some general properties of this function. In
Sec.~\ref{sec:entropy} we introduce a two-parameter generalization of the $BGS$ entropy, and discuss various of its properties. We finally conclude in Sec. 5.

\section{DERIVATION OF THE TWO-PARAMETER GENERALIZATION OF THE LOGARITHMIC FUNCTION}
Eq.~(\ref{eq:fqq1}) constitutes a strongly nonlinear functional equation. A general analytic expression of $\ln_{q,q^\prime} x$ needs to satisfy the limiting cases $\ln_{q,1} x = \ln_{1,q} x = \ln_q x$. 

In order to simplify Eq.~(\ref{eq:fqq1}), we set $x=y$ and, inspired by the expression on the left hand side of
Eq.~(\ref{eq:q_rel3}), write $g(z) = g(\ln_q x ) = \ln_{q,q^{\prime}} x$. By inserting $g(z)$ in
Eq.~(\ref{eq:fqq1}), this functional equation transforms into the much simpler form
\begin{equation}
 g(2z) = 2 g(z) + (1-q^{\prime}) \left[ g(z) \right]^2.
 \label{eq:functional}
\end{equation}
We then verify that the ansatz $g(z) = \frac{1}{1-q^{\prime}} ( b^z-1)$ solves
Eq.~(\ref{eq:functional}) exactly, thus leading to,
\begin{equation}
\ln_{q,q^{\prime}}x = \frac{1}{1-q^{\prime}} \left( b^{\frac{1}{1-q}\left( x^{1-q} -1 \right)}  -1 \right),
  \label{eq:func_gen}
\end{equation}
where $b$ corresponds to a constant that still needs to be determined.
It is important to point out that $\ln_{q,q^\prime}(x)$ solves
Eq.~(\ref{eq:fqq1}) also for the more general case $x \neq y$.

In order to determine $b$, we focus on the point
$\ln_{q,q^{\prime}} 1=0$, which is expected to hold for all $b$ and all $(q,q^{\prime})$, analogously to the fact that $\ln_q 1=0$ for all $q$. We assume also that, at this point, the entire family $\ln_{q,q^{\prime}} x$ has one and the same slope, i.e., 
$\frac{d}{dx}\ln_{q,q^{\prime}} x|_{x=1}=1$, once again in full analogy with $\ln_q x$. Then, we verify that our new
function, Eq.~(\ref{eq:func_gen}), takes the following more specific form: 
\begin{eqnarray}
\ln_{q,q^{\prime}}x =\frac{1}{1-q^{\prime}} \left[ \exp\left( \frac{1-q^{\prime}}{1-q} \left( x^{1-q} -1
    \right) \right) -1 \right].
  \label{eq:sol}
\end{eqnarray}
Indeed, we can prove that $\ln_{q,q^{\prime}} x$ provides a generalized function of the
$q$-logarithm with $b=\exp(1-q^\prime)$ by setting either $q$ or $q^{\prime}$ equal to unity. Eq. (\ref{eq:sol}) can be rewritten as follows:
\begin{equation}
\ln_{q,q^{\prime}}x = \ln_{q^\prime} e^{\ln_q x} \,.
\end{equation}
The relations
\begin{equation}
\ln_{q,1}x = \ln_{1,q}x = \ln_q x 
  \label{eq:limits}
\end{equation}
are easily recovered by evaluating Eq.~(\ref{eq:sol}) in the limits $q
\rightarrow 1$ and $q^{\prime} \rightarrow 1$. 

In Fig.~\ref{fig:fqq} we illustrate the two--parameter
generalized logarithmic function for typical
$q=q^{\prime}$ cases. 
Typical $q\neq q^{\prime}$ cases  are illustrated in Fig.~\ref{fig:f0.3q}.

\section{PROPERTIES}

In this section we present some general properties of the 
two--parameter generalized logarithm, some of which will be useful in the
discussion of Sec.~\ref{sec:entropy}.

\begin{figure}[!h]
  \begin{center}
    \includegraphics[width=0.49\columnwidth]{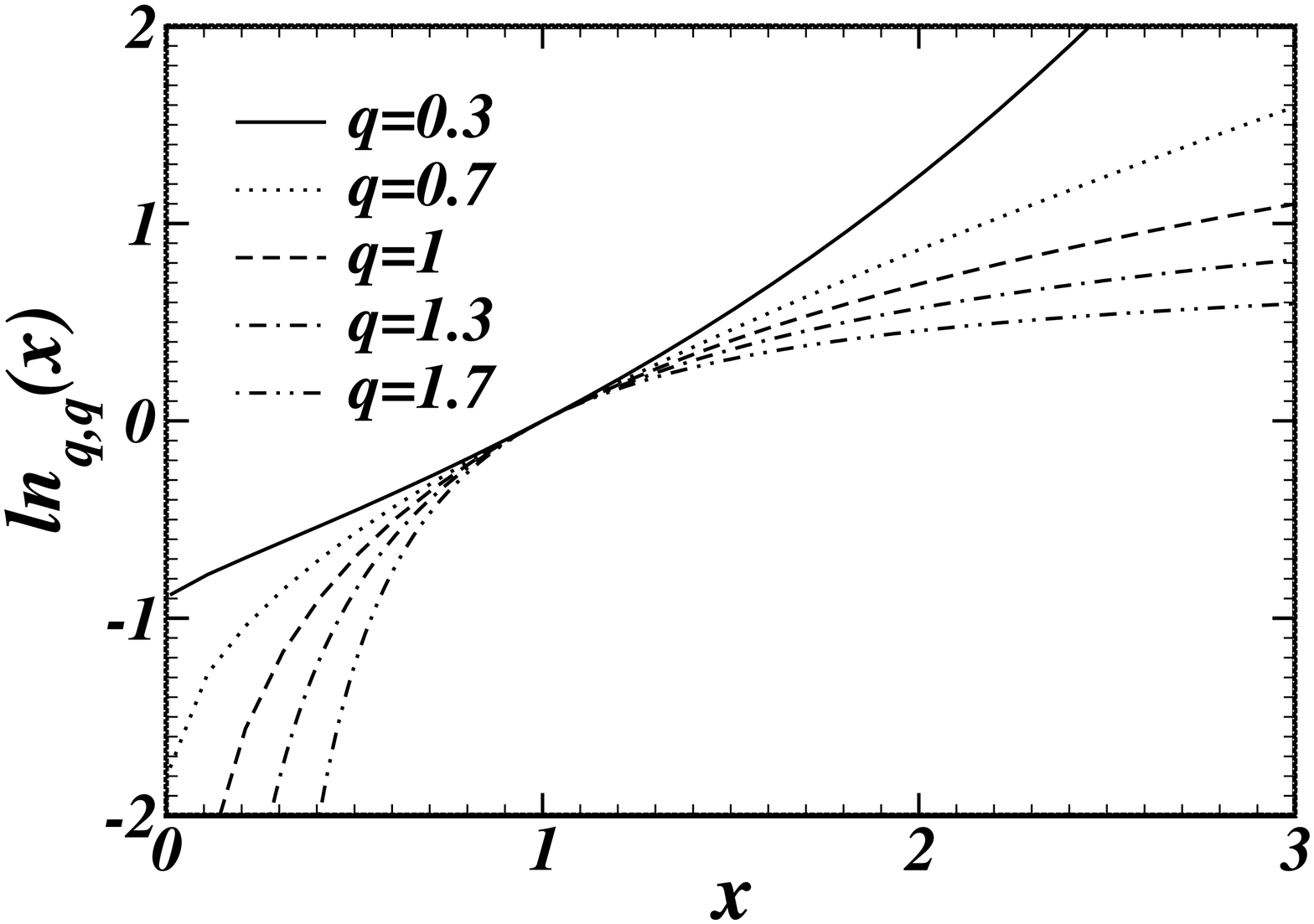}
    \includegraphics[width=0.49\columnwidth]{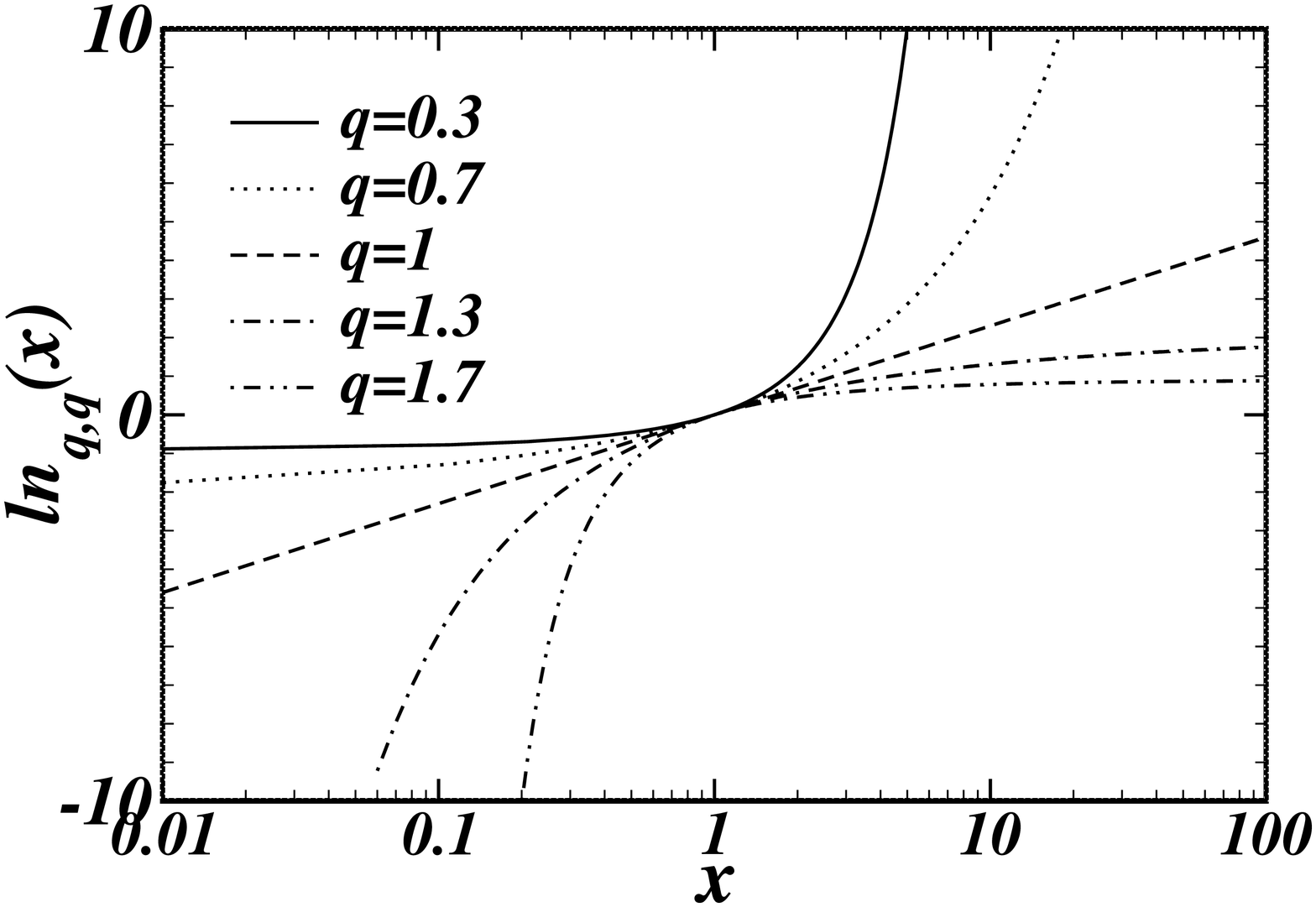}
\caption{Illustration of the generalized logarithm, Eq.~(\ref{eq:sol}),
  setting $q=q^{\prime}$, in linear scales (left) and semi--logarithmic scales (right).
}
    \label{fig:fqq}
\end{center}
\end{figure}

\begin{figure}[!h]
  \begin{center}
    \includegraphics[width=0.49\columnwidth]{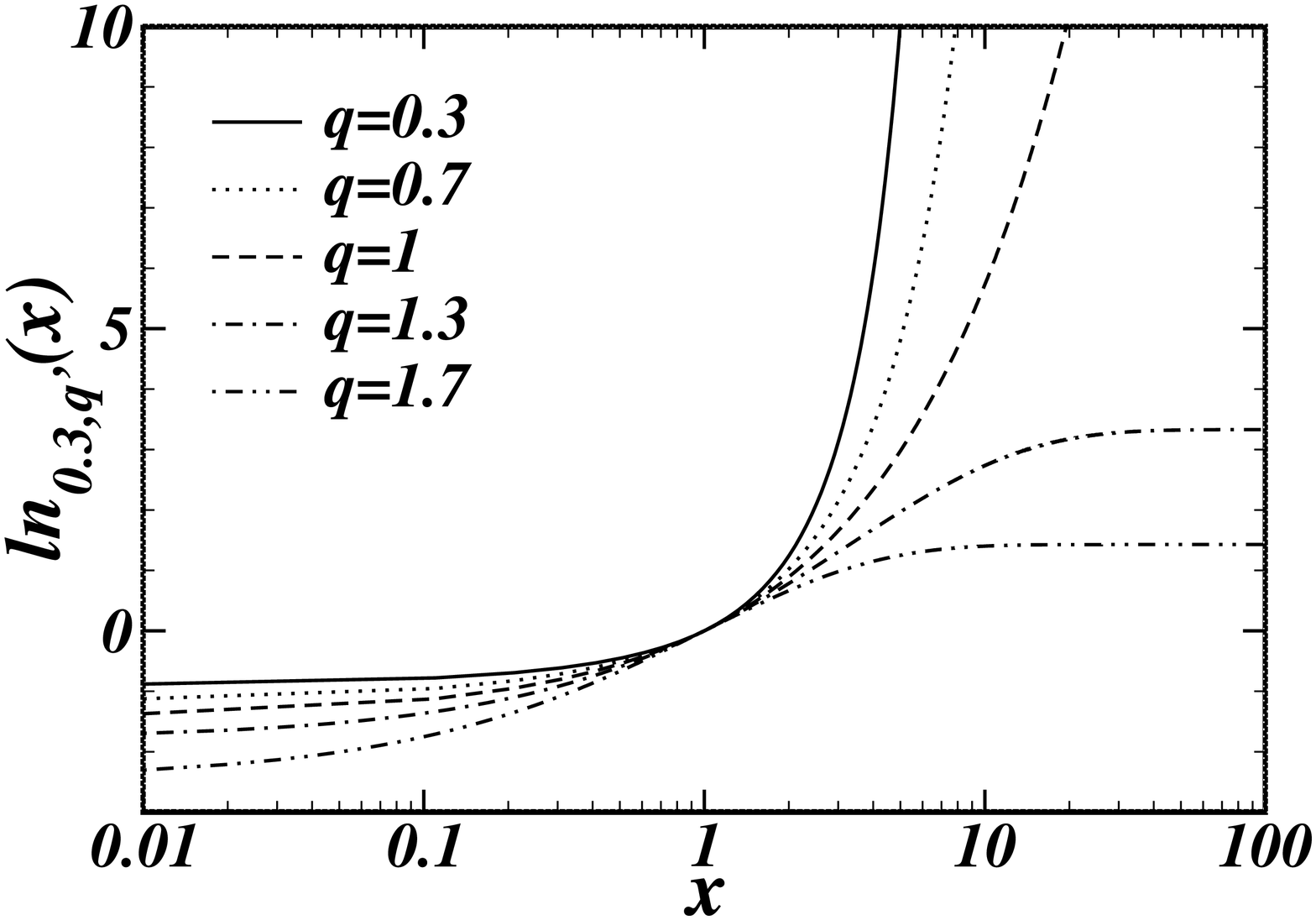}
    \includegraphics[width=0.49\columnwidth]{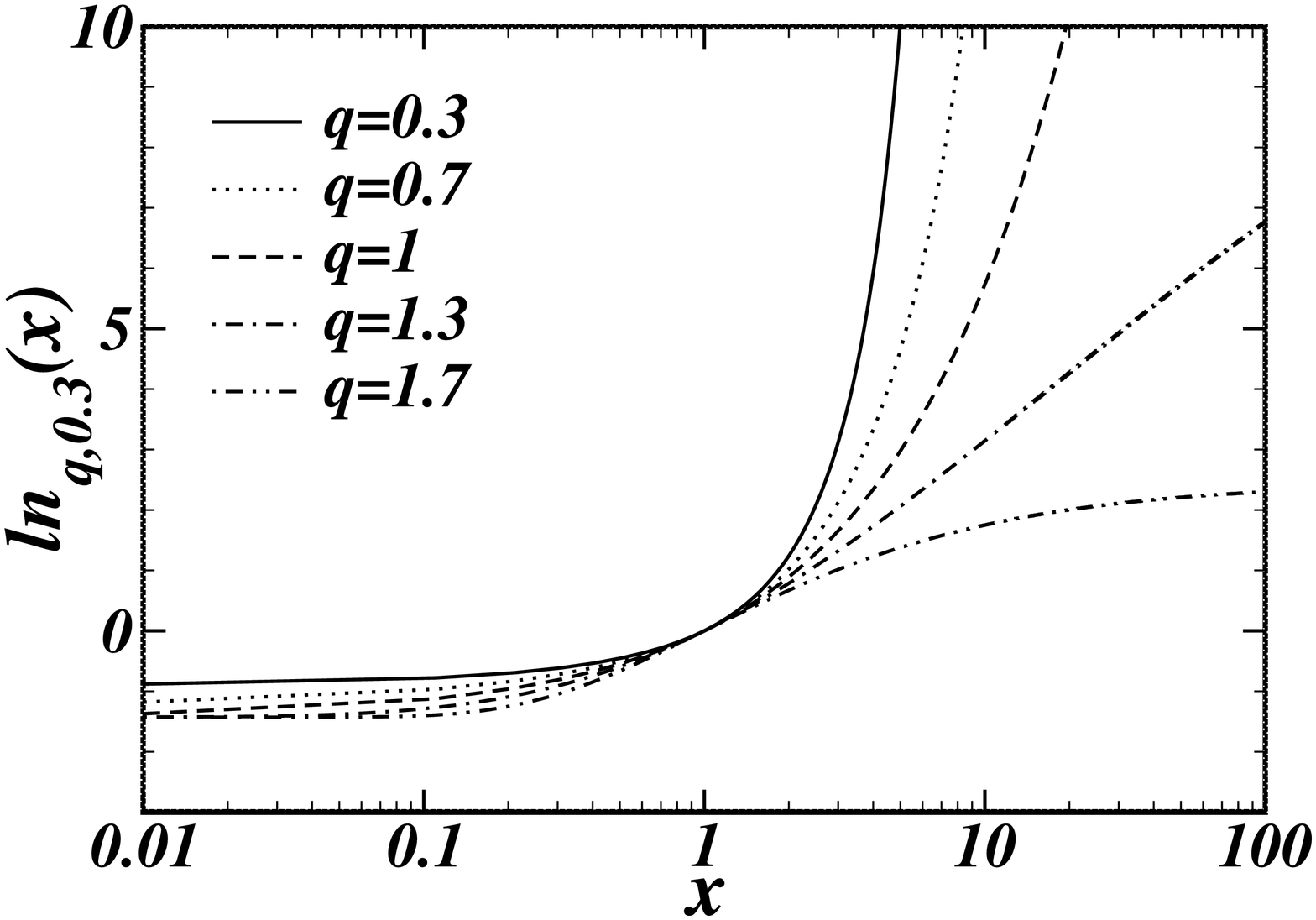}
    \includegraphics[width=0.49\columnwidth]{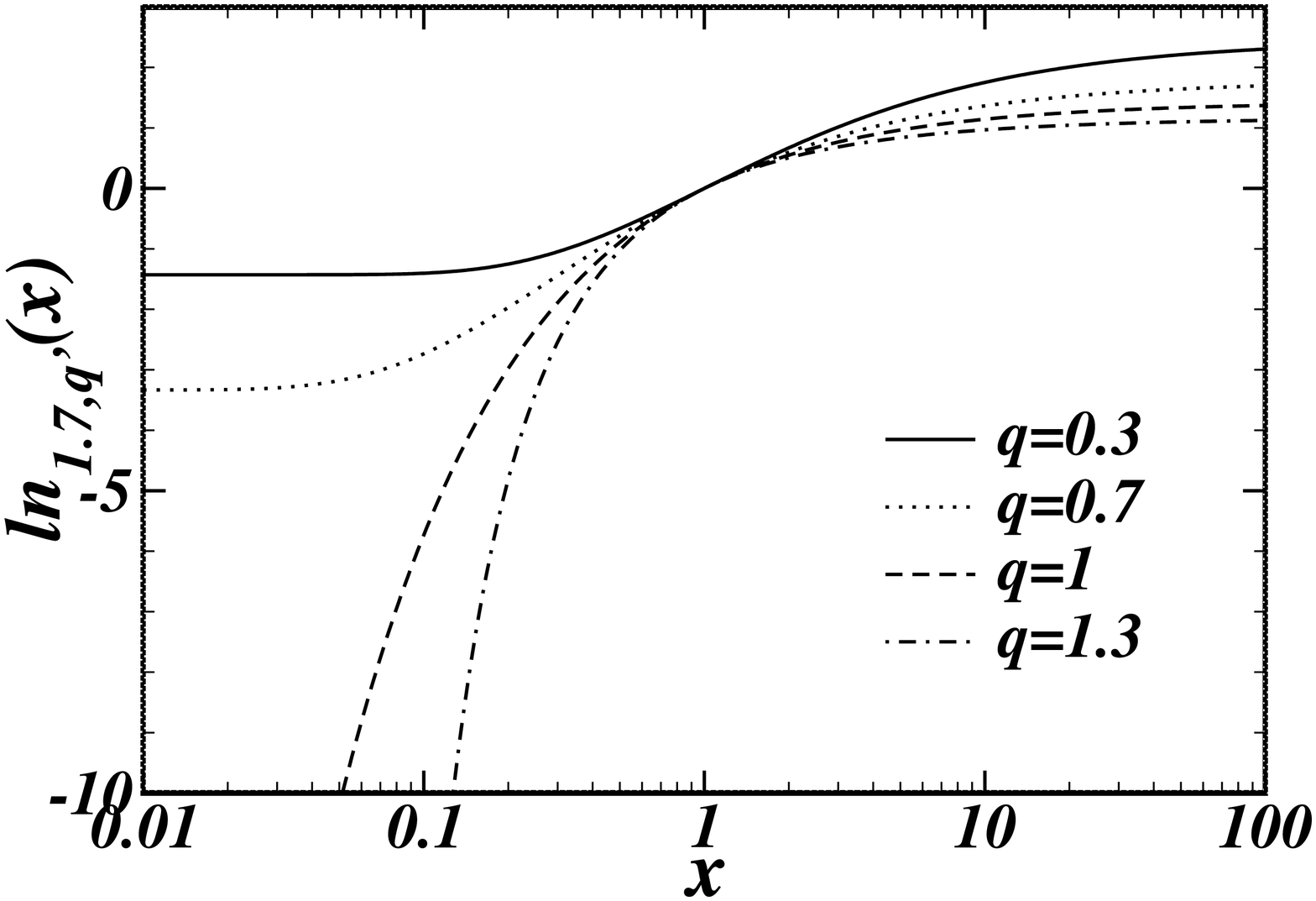}
    \includegraphics[width=0.49\columnwidth]{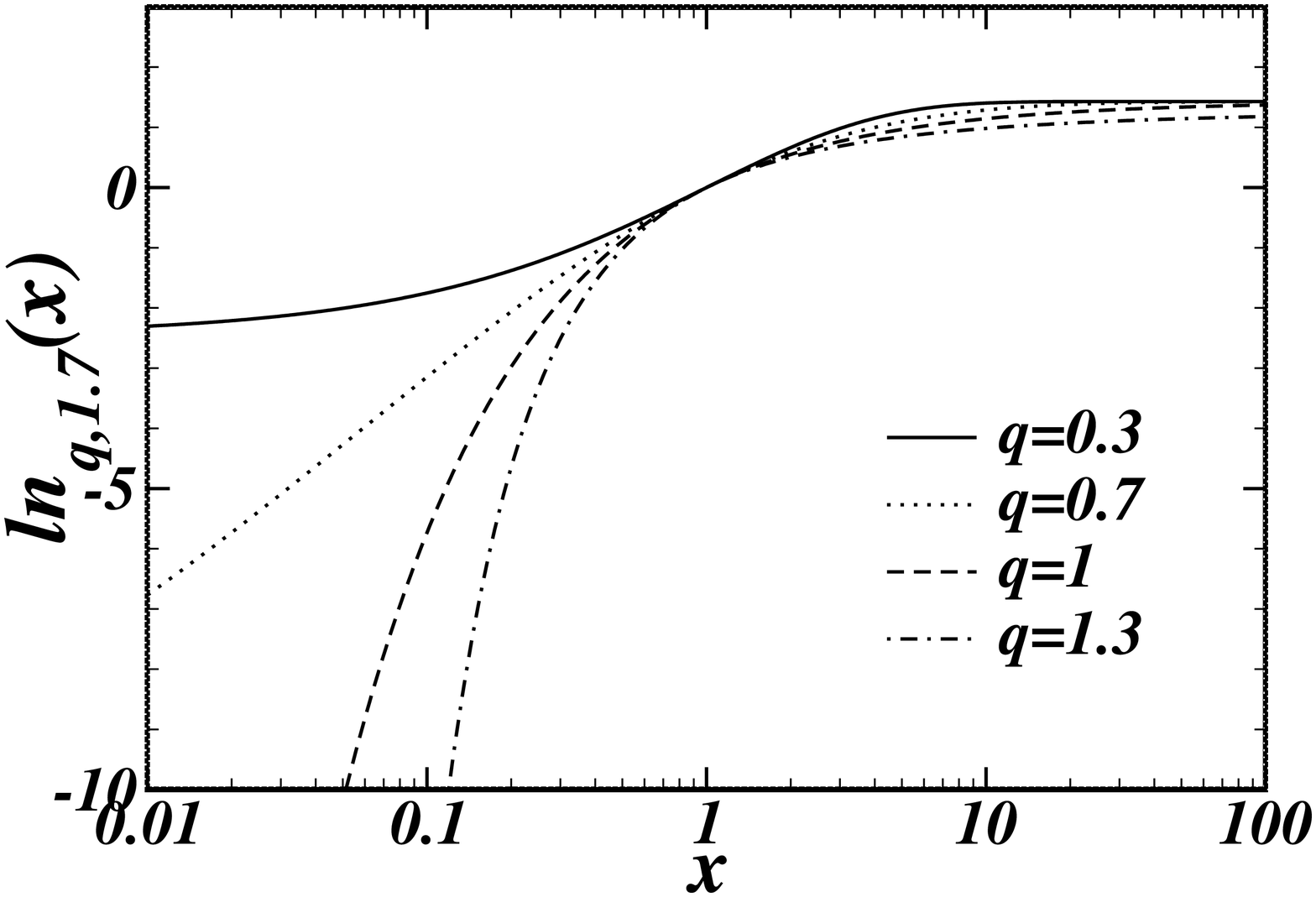}
\caption{Illustration of the generalized logarithm for typical values of the case $q \neq q^{\prime}$. We set
$q=0.3$ and vary $q^{\prime}$ (upper left), then set $q^{\prime}=0.3$ and vary
$q$ (upper right). The lower figures correspond to fixed $q=1.7$ (left) and
$q^{\prime}=1.7$  (right). Note, that the curves in the left figures and the ones in the right figures can be transformed into each other by applying Eq.~(\ref{eq:symm}).}
    \label{fig:f0.3q}
\end{center}
\end{figure}

As can be obtained from Eq.~(\ref{eq:sol}) (and observed in
Figs.~\ref{fig:fqq} and~\ref{fig:f0.3q}), $\ln_{q,q^{\prime}} x$ monotonically increases with $x$.
Thus,  an unique inverse function of $\ln_{q,q^{\prime}} x$, namely the two--parameter generalized
exponential function, exists for arbitrary values of $q$ and $q^{\prime}$:
\begin{equation}
\e_{q,q^{\prime}}^y = \left\{1+ \frac{1-q}{1-q^{\prime}} \ln \bigl[1+ (1-q^{\prime}) y  \bigr] \right\}^{\frac{1}{1-q}} =  e_q^{\ln {e_{q^\prime}}^{y}} \,.
  \label{eq:inverse}
\end{equation}
There are two cut--offs, namely for
$y (1-q^\prime) > 1$ and for $\{...\}<0$. This can be easily 
understood looking at the shape of the generalized logarithm in
Figs.~\ref{fig:fqq} and~\ref{fig:f0.3q}: $\ln_{q,q^\prime} x$ saturates for
$x \rightarrow \infty$ if
at least one of the parameters is greater than unity (either $q>1$ or $q^\prime>1$). Moreover, it also approaches a finite value for $x \rightarrow
0$ if at least one of the parameters is less than unity (either $q<1$ or $q^\prime<1$). 

It is remarkable that the functional form of the present two--parameter generalization of the
logarithmic function contains the standard exponential function,
$\e^x$. The reciprocal happens with the functional form of its inverse function, $\exp_{q,q^\prime} (x)$, which
contains $\ln x$.

Many functions, e.g. $\sin x$ and $\cos x$, can be
expressed in terms of the exponential function. Similarly, it should be possible to construct a
new family of functions using Eqs.~(\ref{eq:sol})
and~(\ref{eq:inverse}). 

Let us calculate here the first derivative of $\ln_{q,q^{\prime}} x$:
\begin{equation}
\frac{d\ln_{q,q^{\prime}}x}{dx} = x^{-q} \exp \left( \frac{1-q^{\prime}}{1-q} \left( x^{1-q} -1 \right) \right).
  \label{eq:deriv}
\end{equation}
The slope of $\ln_{q,q^\prime} x$ remains positive for all $x>0$, which is
consistent with the already mentioned monotonicity of the generalized
logarithm for all values of $(q,q^\prime)$. 

The shape of $\ln_{q,q^{\prime}} x$ in Figs.~\ref{fig:fqq} and~\ref{fig:f0.3q} suggests that
there may exist some symmetry with respect to the point $(x,y)=(1,0)$. Indeed, the 
transformation $q \rightarrow 2-q$, 
well known to be an important relation in nonextensive thermostatistics~\cite{CuradoTsallis:91}, makes
$\ln_{q,q^{\prime}} x$ to satisfy the following property:
\begin{equation}
\ln_{q,q^{\prime}}\frac{1}{x} = -\ln_{2-q,2-q^{\prime}}x.
  \label{eq:symm}
\end{equation}

\section{TWO-PARAMETER GENERALIZATION OF THE BOLTZMANN-GIBBS-SHANNON ENTROPY}
\label{sec:entropy}
We construct here an entropic functional based on the two--parameter generalization of the standard logarithm.

Along the lines of Eqs. (1) and (2) we can construct a new entropic functional as follows: 
\begin{equation}
S_{q,q^{\prime}}  \equiv k \sum \limits_{i=1}^W p_i
\ln_{q,q^{\prime}}\frac{1}{p_i} = \frac{k}{1-q^{\prime}} \sum
\limits_{n=1}^W p_i \left( e^{\frac{1-q^{\prime}}{1-q}\left( p_i^{q-1} 
      -1\right)} -1 \right)  \,,       
\end{equation}
(with $\sum_{i=1}^Wp_i=1$) hence, if $p_i=1/W \,(\forall i)$, $S_{q,q^\prime}=k \ln_{q,q^\prime} W$.

\paragraph{Lesche-stability (or experimental robustness): }
The functional form of $\ln_{q,q^\prime} x$ being analytic in $x$, this entropy is Lesche-stable (experimentally robust)~\cite{CuradoNobre:2004}.

\paragraph{Expansibility: }
A zero-probability ($p_i=0$) state does not contribute
to the entropy if at least one of the parameters is greater than unity ($q>1$
or $q^\prime > 1$).
Note that $q > 0$ ($q^\prime > 0$) in the limit $q^\prime=1$ ($q=1$). 
Therefore $S_{q,q^\prime}(p_1,p_2,...,p_W,0)=S_{q,q^\prime}(p_1,p_2,...,p_W)$,
i.e., $S_{q,q^\prime}$ is expansible within the above given limits
for the two parameters.

\begin{figure}[!h]
  \begin{center}
    \includegraphics[width=0.49\columnwidth]{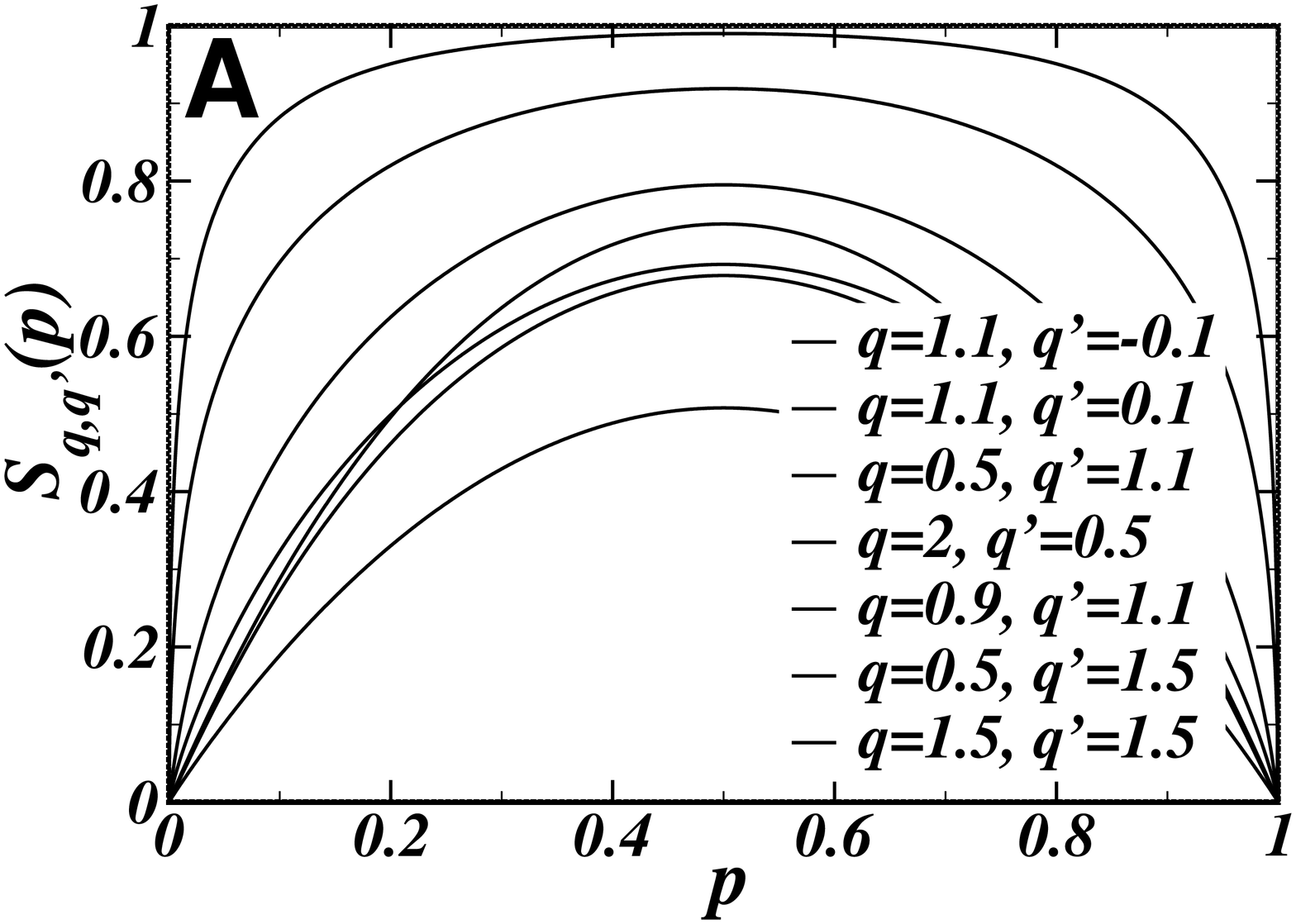}
    \includegraphics[width=0.49\columnwidth]{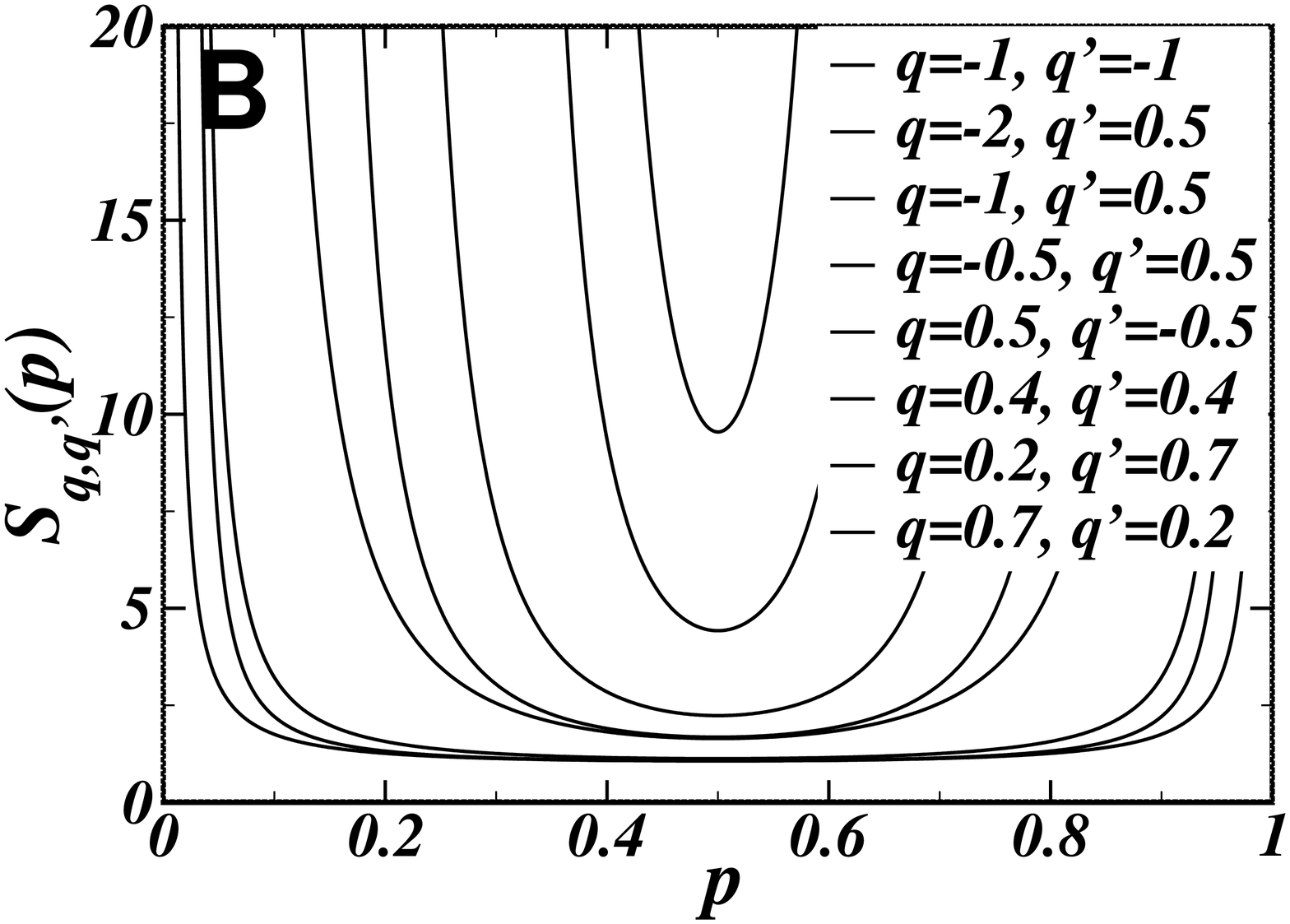}
    \includegraphics[width=0.49\columnwidth]{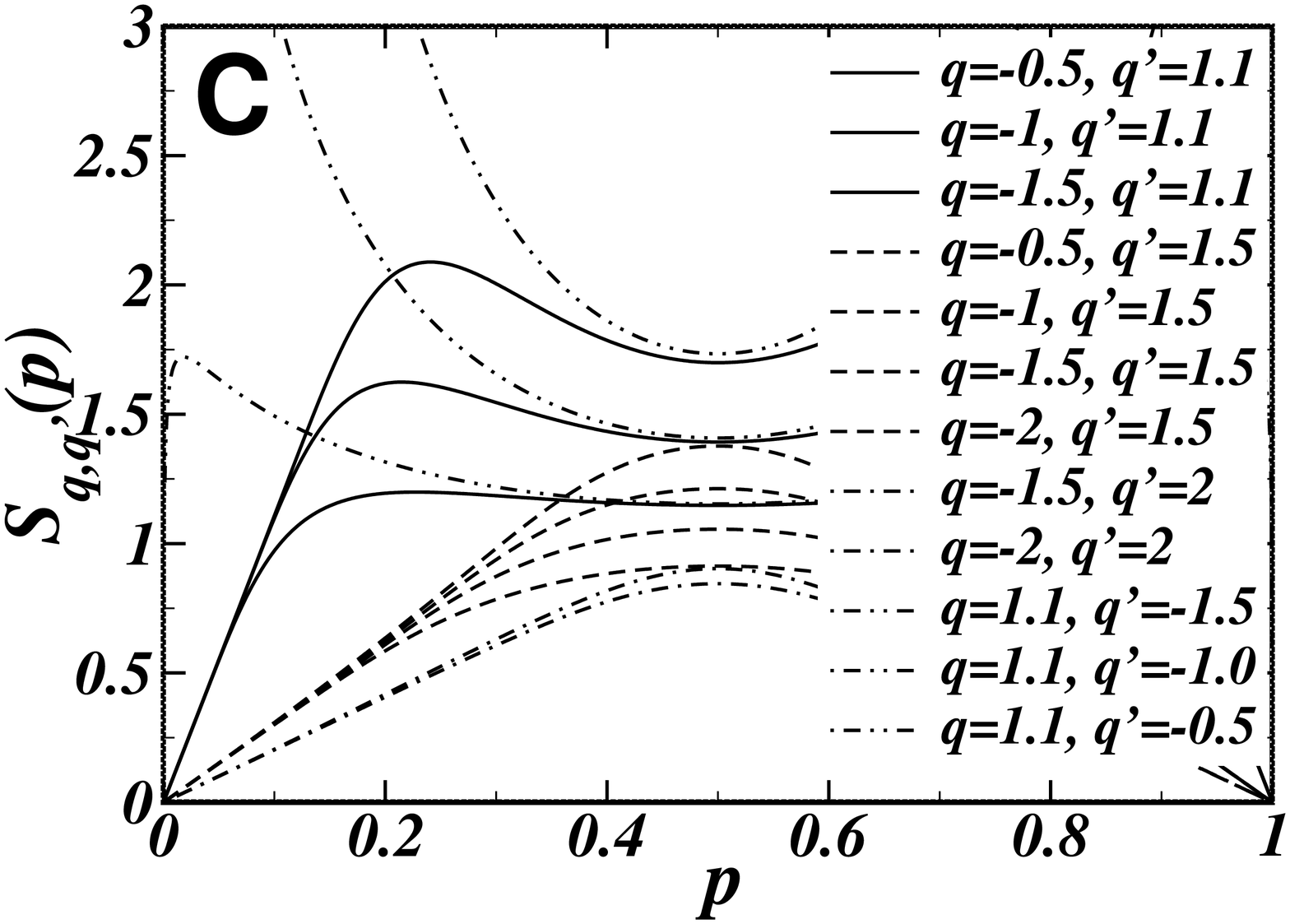}
    \includegraphics[width=0.49\columnwidth]{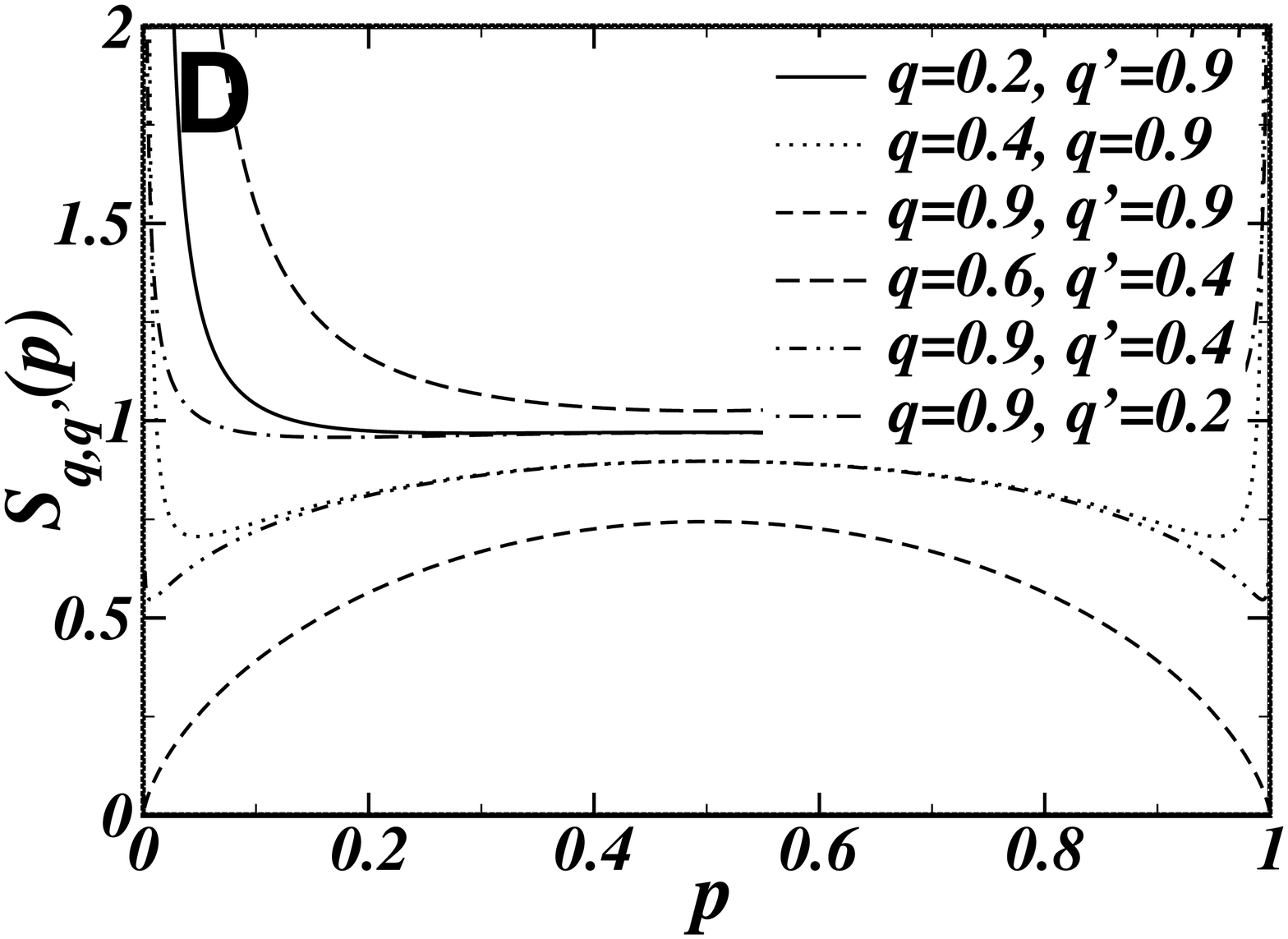}

\caption{Illustration of the two-state generalized entropy $S_{q,q^{\prime}}$ for typical values of $(q,q^{\prime})$. {\bf A}: concavity (region I in Fig.~\protect\ref{fig:diagram}). {\bf B}:
  convexity  (region II in Fig.~\protect\ref{fig:diagram}). {\bf C}: curvature
  can change (region III in Fig.~\protect\ref{fig:diagram}). {\bf D}: curvature
  can change  (region IV in Fig.~\protect\ref{fig:diagram}). }
    \label{fig:entropy2}
\end{center}
\end{figure}

\paragraph{Concavity: } An important requirement for an entropic function is
that its curvature does not change with $\{p_i\}$ (sufficient although not
necessary condition for being concave or convex, i.e.,  
an entropy can still have a fixed sign
of its curvature for parameter values not satisfying this condition). 
Figs.~\ref{fig:entropy2}A-D illustrate $S_{q,q^{\prime}}$ for a system of two
states with probabilities $p$ and $(1-p)$ for typical values of $(q,q^{\prime})$. In Fig.~\ref{fig:entropy2}A
(Fig.~\ref{fig:entropy2}B), the curvature of $S_{q,q^{\prime}}$ is negative (positive) for all
$(q,q^{\prime})$, and the extremum lies at $p=0.5$ as expected. 

In order to determine the regions of the parameter space where the curvature of
$S_{q,q^{\prime}}$ does not change, we focus on the curvature of $p_i
\ln_{q,q^{\prime}}(1/p_i)$. We check whether there is any change of the sign of the curvature in the interval
$0 \le p_i \le 1$.  
We verify  
\begin{equation}
 \frac{d^2}{dp_i^2} p_i \ln_{q,q^{\prime}}\frac{1}{p_i} =  e^{\frac{1-q^{\prime}}{1-q} \left( p_i^{q-1} -1 \right)} \left[ p_i^{2q-3}
  (1-q^{\prime}) - q p_i^{q-2} \right].
\label{eq:2nd_deriv}
\end{equation}
Concavity is guaranteed if Eq.~(\ref{eq:2nd_deriv}) remains negative. 
This is fulfilled if $q+q^{\prime} \geq 1$ in the limit $p_i=1$, and $q^{\prime} \geq 1$ if $q
< 1$ in the limit $p_i \rightarrow 0$. Analogously, convexity is guaranteed if $q+q^{\prime} \leq 1$
in the limit $p_i=1$, and $q,q^{\prime} < 1$ in the limit $p_i \rightarrow 0$. 

\begin{figure}[!h]
  \begin{center}
    \includegraphics[width=0.6\columnwidth]{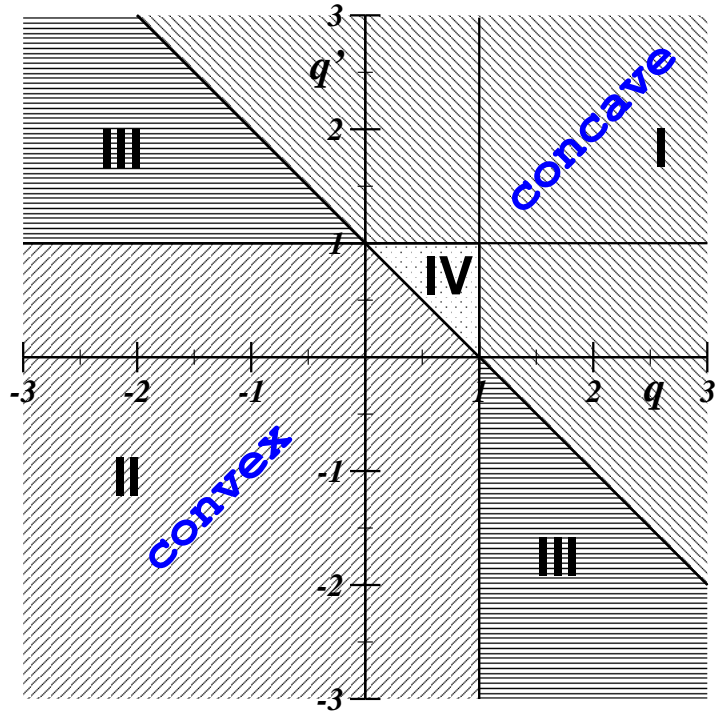}
\caption{The different regions of the entropy
  $S_{q,q^{\prime}}$ with respect to its curvature (see also Fig.~\protect\ref{fig:entropy2}). The point $q=q^\prime=1$ corresponds to the Boltzmann-Gibbs-Shannon entropy. The lines $q=1$ and $q^\prime=1$ correspond to the nonadditive entropy $S_q$. This diagram exhibits $(q,q^\prime)\rightarrow (q^\prime,q)$ symmetry although generically $S_{q,q^\prime} \ne S_{q^\prime,q}$.}
    \label{fig:diagram}
\end{center}
\end{figure}

Fig.~\ref{fig:diagram} exhibits the various regions for the curvature with respect to $q$ and $q^{\prime}$. We
can see that two regions (III and IV) may have non fixed curvature sign. An
analysis of the shape of $S_{q,q^{\prime}}$ with $W=2$ for several values of
$q$ and $q^{\prime}$ confirms this but suggests
that parts of regions III and IV may still have fixed curvature sign (see also
Fig.~\ref{fig:entropy2} for examples of $S_{q,q^{\prime}}$ corresponding to
these regions).


\paragraph{About composability: } Consider a system $A+B$ composed by two probabilistically {\it independent} subsystems $A$ and $B$ (i.e., $p^{A+B}_{i,j}=p^A_i p^B_j$). An entropic functional $S$ is said to be {\it composable} if its value $S(A+B)$ depends from the probabilities $\{p^A_i\}$ $\{p^B_j\}$ {\it only through} $S(A)$, $S(B)$ and generic indices, i.e., if $S(A+B)=\Phi[S(A),S(B),indices]$, where $\Phi$ is some univalued function.   Any {\it additive} entropic functional (such as the BGS and the Renyi ones) trivially is composable (e.g., $S_{BGS}(A+B)=S_{BGS}(A)+S_{BGS}(B)$). The {\it nonadditive} entropy $S_q$ ($q \ne 1$) also is composable since $\frac{S_q^{A+B}}{k}=\frac{S_q^A}{k} \oplus_q \frac{S_q^B}{k}=\frac{S_q(A)}{k}+\frac{S_q(B)}{k}+(1-q)\frac{S_q(A)}{k}\frac{S_q(B)}{k}$. Composability is an important property for thermodynamical applications, not so for applications such as signal or image processing, cybernetics, and others.

Let us now check whether $S_{q,q^\prime}$ is composable. We start by considering a micro--canonical ensemble, i.e., equal probabilities. We have
$S^{A+B}_{q,q^{\prime}}  = k \ln_{q,q^{\prime}}(W_A W_B)$, $S^{A}_{q,q^{\prime}}  = k \ln_{q,q^{\prime}}W_A$ and $S^{B}_{q,q^{\prime}}  = k \ln_{q,q^{\prime}}W_B$. It straightforwardly follows that
\begin{eqnarray}
& \frac{S^{A+B}_{q,q^{\prime}}}{k} = \frac{1}{1-q^\prime} \Bigg\{  e^{ \frac{1-q^{\prime}}{1-q} \ln \bigg[1+
     (1-q^{\prime}) \frac{S_{q,q^\prime}^A}{k}  \bigg] 
\ln \bigg[1+ (1-q^{\prime}) \frac{S_{q,q^\prime}}{k}^B \bigg]} & \nonumber \\
& \qquad \qquad  \times \bigg[1+
   (1-q^{\prime}) \frac{S_{q,q^\prime}^A}{k}  \bigg] \bigg[1+ (1-q^{\prime})
   \frac{S_{q,q^\prime}^B}{k} \bigg] - 1 \Bigg\} ~.
\label{eq:additivity1}
\end{eqnarray}
\noindent
The meaning of such a relation ({\it pseudo-additivity}) is uneasy to interpret. Nevertheless, we can present
Eq.~(\ref{eq:additivity1}) in a simpler manner by introducing the transformation $Y(S)\equiv \ln \left\{1+
  \frac{1-q}{1-q^{\prime}} \ln \left[ 1+(1-q^{\prime}) (S/k)  \right]  \right\}$. We then obtain 
\begin{equation}
Y(S^{A+B}) = Y(S^{A}) +  Y(S^{B}).
\label{eq:comp_simpl}
\end{equation}
Composability of the present entropic form $S_{q,q^\prime}$ requires that the same relation
in Eq.~(\ref{eq:comp_simpl}) remains valid for other thermostatistical
statistics such as, for instance, that of the canonical ensemble. 
However, our studies indicate that this is {\it not} the case for arbitrary distributions
$\{p_i\}$. Numerical verifications confirmed this result. Therefore $S_{q,q^\prime}$ is in general not composable.

\section{CONCLUSION}

We found an explicit generalized logarithmic function which nicely closes the picture obtained with the $q$-logarithm using the
$q$--sum and the $q$--product as basic mathematical operations. It is 
possible to use the new logarithm,
$\ln_{q,q^{\prime}} x$, and its inverse function, $e_{q,q^\prime}^x$, in order to construct a new family
of functions such as $\sin_{q,q^{\prime}} x$ and $\cos_{q,q^{\prime}} x$. It remains to
be tested if these new findings can be useful to describe, in some way, some behavior of natural, artificial or social complex systems, or if they can be useful in applied areas such as signal and image processing, cybernetics, optimization algorithms, information theory, among others. Even if they do not immediately exhibit applications, 
they remain as elegant mathematical constructions. Does $\ln_{q,q^{\prime}} x$ constitute a
``super--generalized'' logarithm possibly useful as the basis for a more general
entropy? We engaged this question by showing that various important properties (concavity, Lesche-stability, expansibility) presumably needed for a physical entropy are fulfilled in this case providing that some minimal restrictions are imposed to
$(q,q^{\prime})$. However, one crucial property failed to be valid in general, namely composability. The
fact that this entropy is not composable leads to the assumption that
it may be not able to play a thermodynamical role comparable to the nonadditive entropy $S_q$ on which nonextensive statistical mechanics is based.

As a last remark, let us mention that, along the present lines, Renyi' s entropy $S_q^R \equiv \frac{\ln \sum_{i=1}^W p_i^q}{1-q}$ can also be (trivially) generalized into a two-parameter one, namely $\frac{\ln_{q^\prime} \sum_{i=1}^W p_i^q}{1-q}$ . 

\section*{Acknowledgements}
One of us (C. T.) has benefited from interesting conversations with E.P. Borges, M. Gell-Mann and R. Hersh at the Santa Fe Institute, New Mexico, where this project started. Financial support by Pronex/MCT, CNPq and Faperj (Brazilian Agencies) is gratefully acknowledged as well.\\

\bibliographystyle{unsrt}
\bibliography{entropy}

\end{document}